\documentstyle[12pt]{article}

\begin{document}

\begin{center}

{\Large  {\bf  Pseudo-Hermitian Interactions in Dirac Theory: Examples }}

\vspace{.5in}

 Bhabani Prasad Mandal \footnote{bhabani.mandal@gmail.com ; bhabani@bhu.ac.in}
and Saurabh Gupta \footnote{guptasaurabh4u@gmail.com}

Department of Physics, Banaras Hindu University, Varanasi-221005,
INDIA.

 \vspace{.9in}

{\bf Abstract}
\end{center}

We consider a couple of examples to study the  pseudo-Hermitian interaction 
in relativistic quantum
 mechanics. 
Rasbha interaction, commonly used to study the spin Hall effect, is considered
with imaginary coupling. The corresponding Dirac Hamiltonian is shown to be
parity pseudo-Hermitian. In the other example we consider parity 
pseudo-Hermitian scalar interaction with arbitrary parameter in Dirac theory. 
In both the cases we show that  the
energy spectrum is  real and all the  other features of
non-relativistic  pseudo-Hermitian  formulation are present.
Using the spectral method the positive definite metric operator ($\eta$)
has been calculated explicitly  for both the models to ensure positive definite
norms for the state vectors.

\newpage

\vspace{.5in}
\section{Introduction}

\renewcommand{\theequation}{1.{\arabic{equation}}}
\setcounter{equation}{0}

Non-Hermitian quantum mechanics has recently created a lot of interest.
 This is due to the observation by Bender and Boettcher \cite{ben,ben11}
that with properly defined boundary conditions, the spectrum of the
system described by the Hamiltonian $ H= p^2 + x^2 (ix)^\nu, \  \
\nu \geq 0$ is real positive and discrete. The reality of the
spectrum is a consequence of unbroken PT (combined Parity [P] \&
Time reversal [T]) invariance of the Hamiltonian, $ [H,PT]\psi
=0$ and the spectrum becomes partially complex when the PT
symmetry is broken spontaneously \cite{pt3,pt31}.

This new result has given rise to growing interest in the
literature, see for examples, \cite{pt3}-\cite{co}. Past few years many non-Hermitian but PT symmetric
systems have been investigated including (i)  quantum mechanics of single particle in
one space dimension \cite{ben}-\cite{kha}, (ii) exactly solvable many particle 
quantum systems
in one space dimension \cite{bm}-\cite{bn}, (iii) field theoretic models \cite{ft}-\cite{conf}. To develop a consistent quantum theory for these
systems, one encounters severe difficulties \cite{wei,jap}.  Particularly,
 the
eigenstates of PT symmetric non-Hermitian Hamiltonians, with real
eigenvalues only, do not satisfy standard completeness relations.
 More importantly  the eigenstates have negative norms if one
takes the natural inner product associated with such systems
defined as \begin{equation}  <f\mid g>_{PT} = \int d^3 x [PT f(x)] g(x).
\end{equation}  These
problems are overcome by introducing a new symmetry [C], analogous
to charge conjugation symmetry,  associated with all such systems
with equal number of negative and positive norm states \cite{pt1,co}. This
allows to introduce an inner product structure associated with CPT
conjugate as
\begin{equation}
  <f\mid g>_{CPT} = \int d^3 x [CPT f(x)] g(x),
  \label{cpt}
  \end{equation}
for which the norms of the quantum states are positive
definite and one gets usual completeness relation.
 As a result,
the Hamiltonian and its eigenstates can be extended to complex
domain so that the associated eigenvalues are real and underlying
dynamics is unitary. Thus we have a fully consistent quantum theory
for non-Hermitian but PT invariant systems.

In an alternative approach \cite{alir}-\cite{gey}, it has been  shown  that the
reality of the spectrum of a non-Hermitian system is due to so called
pseudo-Hermiticity properties of the Hamiltonian. A Hamiltonian
is called $\eta $ pseudo-Hermitian if it satisfies the relation 
\begin{equation}
\eta H \eta ^{-1} = H^\dagger\label{ps},
\end{equation}  where $\eta $ is some 
linear Hermitian and invertible operator. All PT symmetric non-Hermitian systems
are pseudo-Hermitian where parity operator plays the role of
$\eta $. However there are many pseudo-Hermitian
systems with completely real spectrum which are not PT symmetric \cite{bpm}-\cite{yb}.
In the formulation of pseudo-Hermitian quantum mechanics, the
scalar product is defined as
 \begin{equation} <f\mid g>_\eta =
<f\mid\eta g>=\int d^3 x f^*\eta g \label{eta},
\end{equation}
to get rid of the conceptual difficulties arise in such theories. 
 Equation (\ref{cpt}) which defines the scalar product  in case of non-Hermitian
PT invariant theory is a special case of Eq. (\ref{eta}) \cite{alin}.

If the Hamiltonian is diagonalizable and has a real and discrete spectrum, 
it is always possible to find a positive definite $\eta$ \cite{alir,rev0}
 such that the norms of all the eigenstates
 become positive definite and they satisfy usual
completeness relation. Thus one can  have completely consistent quantum
theory for the pseudo-Hermitian systems provided atleast one positive definite
$\eta $ is contructed.

 However,  most of the previous works in the  pseudo-Hermitian quantum
mechanics have been carried out in the non-relativistic framework \footnote{
Some  attempts to study relativistic PT invariant non-Hermitian system can
be found in Refs. \cite{rev0}-\cite{rev}.}.
The purpose of this paper is to extend the results of
pseudo-Hermitian quantum mechanics for  relativistic systems. Here we
 consider a couple of examples of pseudo-Hermitian
Hamiltonian and show that the energy spectrum obtained by solving
corresponding Dirac equation is also real. In the first example, we 
study the famous Rashba  interaction \footnote{ This type
of spin orbit coupling is widely used  to study spin Hall effect \cite{she,she11,she12,she13}. In
the simplest version of spin Hall effect, an electric current passes through
a sample with spin orbit interaction and induces a spin
polarization near the lateral edges, with opposite polarization  at opposite edges.
 This effect does not require an external magnetic
field or magnetic order in the equilibrium state before the current is
applied. } \cite{ras} with imaginary coupling which is
pseudo-Hermitian with respect to parity operator. We consider a Dirac
particle in 1+1 dimension, moving under arbitrary scalar pseudo-Hermitian
interaction in other example.
 The norms of the state vectors, defined according to the modified
rule of scalar product [ i.e. Eq. (\ref{eta})] will be positive definite
 if we find a positive $\eta $ for the above two models. We have constructed
positive definite $\eta $  for both the models using spectral method \cite{spec,spec1}.

This paper is arranged as follows. In section II,
we consider the  motion of a Dirac particle in the x-y plane experiencing
Rashba interaction with imaginary coupling. The pseudo-Hermitian
scalar interaction of Dirac particle in 1+1 dimension is
discussed in section III.  Positive definite 
$\eta $ for both the models have been constructed explicitly in section IV.
 Last
section reveals for conclusion and summary.

\section{ Dirac equation with Rashba interaction}

\renewcommand{\theequation}{2.{\arabic{equation}}}
\setcounter{equation}{0}

We consider the motion of a particle in the x-y plane described by
the Hamiltonian
\begin{equation}
H= c\mbox{\boldmath $\alpha $} \cdot {\mathbf p } +\beta m_0 c^2 +\lambda_1
(\mbox{\boldmath $\sigma $}\times{\mathbf p})\cdot {\mathbf\hat{z}} \label{2.1}
,\end{equation}
where the last term (with real $\lambda_1$) is known as the  Rashba
interaction term. This type of interaction is  widely used  to
discuss spin Hall effect \cite{she,she11,she12,she13} in non-relativistic formulation.
However, the Hamiltonian is no longer Hermitian if we consider the
$\lambda_1 $ is complex.
 Further it
can be shown that this Hamiltonian is a parity pseudo-Hermitian
when $\lambda_1 (\equiv i\lambda)$ is purely imaginary.
\begin{eqnarray}
H^\dag &=& c\mbox{\boldmath $ \alpha $}\cdot {\mathbf p} +\beta m_0 c^2
-i\lambda (\mbox{\boldmath $\sigma $}\times{\mathbf p})\cdot{\mathbf \hat{z}} \nonumber
\\ &\neq & H \nonumber  \\ &=& P H P^{-1},
\end{eqnarray}
where $P (=\beta e^{i\delta})$ is parity operator in Dirac theory.
({\boldmath $\sigma $} $\times {\mathbf p} )\cdot {\mathbf \hat{z}}$ changes sign under parity
transformation $$ P(\mbox{\boldmath $\sigma $}\times {\mathbf p} )\cdot {\mathbf \hat{z}}P^{-1} =
-(\mbox{\boldmath $\sigma $}\times {\mathbf p} )\cdot {\mathbf \hat{z}}, $$ as under parity:
  
{\boldmath $\sigma $}$\rightarrow ${\boldmath $\sigma $} ,$ \ {\mathbf p}\rightarrow {\mathbf -p}, \ {\mathbf \hat{z}}\rightarrow
-{\mathbf \hat{z}} $  and  $\beta$ anti-commutes with $\sigma_x , \sigma_y$.

Now we  solve the Dirac equation $H\psi =E\psi $ for this
system to find the energy eigenvalues. To do  that we  consider
the wave function in terms of  components, $ \psi
=\left  (\begin{array}{c} \phi    \\  \chi
\end{array} \right ) $
 and restrict ourselves to 2-dimension only.
Further, we consider the particular representation of Dirac
matrices in 2-dimension as, $ {\mathbf\alpha}={\mathbf\sigma} ,
\beta =\left (\begin{array}{cc} 1 & 0   \\  0 &-1
\end{array}\right ),$
 where $ {\mathbf\sigma_i} [\mbox{ for } i=x,y,z]$ are Pauli spin
matrices. Then the Hamiltonian in equation (\ref{2.1}) can be written
as
\begin{eqnarray}
H&=& c\sigma _x  p_x + c\sigma_y p_y +\beta m_0c^2 + i
\lambda(\sigma_xp_y -\sigma_y p_x) \nonumber  \\ &=& \left [
\begin{array}{cc} m_0 c^2 & (c-\lambda)p_-  \\
(c+\lambda)p_+ & -m_0c^2 \end{array} \right ], \label{2.2}
\end{eqnarray}
where $ p_{\pm} = p_x\pm ip_y. $ The Dirac equation  $H\psi =
E\psi $ can be written as pair of coupled equations of the
components $\psi $ and $\chi$ as
\begin{eqnarray}
(c-\lambda)p_- \chi &=& (E-m_0c^2)\phi , \nonumber\\ (c+\lambda)p_+
\phi &=& (E+m_0c^2)\chi \label{2.3}.
\end{eqnarray}
On eliminating one component $\chi$ in terms of others, we obtain
the equation
\begin{equation}
p_-p_+\phi = \epsilon\phi \label{2.4},
\end{equation}
where $\epsilon =\frac{E^2-m_0^2c^4}{c^2-\lambda^2}$.  Equation
(\ref{2.4}) can be interpreted as Schrodinger equation for a free
particle moving in the x-y plane.  The corresponding energy
eigenvalues $\epsilon$ are given as $\hbar^2(k_x^2 +k_y^2)$, which
are real and positive. Therefore the energy eigenvalues E for
the Dirac Hamiltonian for the pseudo-Hermitian system, described in
Eq.(\ref{2.1}), are given by
\begin{equation}
E=\pm \sqrt{m_0^2 c^4 +(c^2-\lambda^2)\hbar^2(k_x^2+ k_y^2)}
\label{2.5}.
\end{equation}
Thus the energy spectrum for a pseudo-Hermitian [ Rashba interaction
with imaginary coupling ] relativistic system is completely real for
$\lambda^2 < c^2 $.
Same conclusion can be drawn by eliminating other component $\phi
$ from the Eqs. in (\ref{2.3}).

\section{Dirac equation with scalar pseudo-Hermitian interaction}

\renewcommand{\theequation}{3.{\arabic{equation}}}
\setcounter{equation}{0}

Let us consider a Dirac particle of rest mass $m_0$  subjected
to a scalar pseudo-Hermitian potential  $V_s$. The dynamics of
such a system can be described by the Hamiltonian
\begin{equation}
H= c\mbox{\boldmath $\alpha $}\cdot{\mathbf p} +\beta m_0 c^2 +V_s .\label{3.1}
\end{equation}
For simplicity, we restrict ourselves to one space dimension and
we choose the scalar pseudo-Hermitian potential as
\begin{equation}
V_s = V(x) \left ( \begin{tabular}{c c} 0&1  \\  -1 &0
\end{tabular} \right ),
\label{3.2}
\end{equation}
where $V(x)$ is an arbitrary real function and assumed to have even parity,
 $V(-x) = V(x)$, for later convenience. We note that the
Hamiltonian given in Eq. (\ref{3.1}) is not Hermitian as $V_s^\dag =
-V_s \neq V_s$, but it is parity pseudo-Hermitian, i.e.
\begin{equation}
H^\dag = PH P^{-1},
\end{equation}
where $P$ is the parity operator and in relativistic formulation
is given by  $P= \beta e^{i\delta}$. This is so because, $
c\alpha\cdot {\mathbf p} +\beta m_0 c^2$ is a parity invariant term and
$$PV_s P^{-1} = \beta e^{i\delta} V(x)\left ( \begin{tabular}{c c}
0&1  \\  -1 & 0
\end{tabular} \right )\beta^{-1}e^{-i\delta}= -V_s = V_s^\dag ,$$
where we have chosen a particular representation of Dirac matrices
$\alpha$ and $\beta$ in 1+1 dimension as $$ \alpha_x = \sigma_x
\mbox{ and }\beta= \left ( \begin{tabular}{c c} 1&0  \\  0 &-1
\end{tabular} \right ), $$  $\sigma_x$ is the first of Pauli's $ 2\times 2$ spin
matrices. 
1+1 dimensional Dirac equation for the system can be written as
\begin{equation}
\left [ c\sigma_x p_x +\beta m_0 c^2 +V_s\right ]\psi= E\psi .
\label{3.3}
\end{equation}
By considering $\psi=\left ( \begin{array}{c} \phi  \\  \chi
\end{array} \right ),$ the above equation can be written in
component form
\begin{eqnarray}
\left [ V(x)- ic\hbar\frac{d}{dx} \right ] \chi &=& (E-m_0c^2)\phi ,
\label{3.35}
\\ \left [- V(x)- ic\hbar\frac{d}{dx} \right ] \phi &=& (E+m_0c^2)\chi
\label{3.4}.
\end{eqnarray}
By eliminating $\chi$ from the above coupled differential
equations, we obtain
\begin{equation}
\left [-\hbar^2\frac{d^2}{dx^2}+U \right ]\phi =\epsilon \phi
\label{sch},
\end{equation}
where $$ U= -i\hbar\frac{dV}{dx}-V^2 \mbox{ and } \epsilon=
E^2-m_0^2c^4 .$$  Eq. (\ref{sch}) is nothing but non-relativistic
Schrodinger equation for a particle of mass $m=\frac{1}{2}$
subjected to a  complex potential $U$ and $\epsilon $ is the
energy eigenvalues for the particle. 
Even though  the potential U
is complex, remarkably it is parity pseudo-Hermitian as $$U^* = \tilde{P}U\tilde{P}^{-1}.$$ 
where $\tilde{P}$ is parity operator in non-relativistic quantum theory.
Note we have assumed $V(-x) = V(x) $. Hence  the energy eigenvalues,$\epsilon$
for this effective non-relativistic theory is 
real or occure in complex conjugate pairs\cite{ali}depending on whether
the symmetry is broken or not.
. The energy eigenvalues of the relativistic particle is given
in terms of $\epsilon$ by
\begin{equation}
E= \pm\sqrt{\epsilon+m_0^2c^4} \label{e}.
\end{equation}
Now if $\epsilon$ is real and positive, the whole spectrum of the
relativistic particle is real without any restriction. On the
other hand, if $\epsilon$ is real but negative  i.e. $\epsilon=
-|\epsilon|$, then also the spectrum is real provided
$|\epsilon|<m_0^2c^4 $. 
However for arbitrary $V(x)$ energy  eigenvalues are not real always
as expected.
Exactly similar conclusion can be drawn by
eliminating $\phi$ from the equations (\ref{3.35}) and (\ref{3.4}).

Now we consider a special case when, $V(x)= V_0$, independent of x

\begin{equation}
V_s = V_0 \left ( \begin{tabular}{c c} 0&1  \\  -1 &0
\end{tabular} \right ),
\label{3.2}
\end{equation}
where $V_0$ is an arbitrary real constant. The Dirac equation in Eq. \ref(3.1) 
 can be written as

\begin{eqnarray}
\left [ \begin{array}{cc} m_0c^2 & cp_x+V_0  \\  cp_x-V_0 &-m_0c^2
\end{array} \right ] \ \left ( \begin{array}{c} \phi \\ \chi \end{array} \right )
= E\left ( \begin{array}{c} \phi \\ \chi \end{array} \right ),
\label{3.4}
\end{eqnarray}
and in terms of components,
\begin{eqnarray}
(cp_x+V_0)\chi &=& (E-m_0c^2)\phi , \nonumber \\
(cp_x-V_0)\phi &=& (E+m_0c^2)\chi .
\end{eqnarray}

By solving  the above coupled differential
equations, we obtain the energy eigenvalues as 
\begin{equation}
E= \pm\sqrt{\hbar^2 c^2 k_x^2 +m^2_0c^4-V_0^2}
\label{3.7}
\end{equation}
with \begin{equation}
p_x^2\phi = \hbar^2k_x^2\phi
\end{equation}

The energy eigenvalues are always real  for sufficiently weak pseudo-Hermitian interaction.

\section{ Construction of positive definite $\eta $}

We have shown in both the examples that even in relativistic quantum mechanics we can have
real eigenvalues if the relativistic Hamiltonian is parity pseudo-Hermitian. In order to have
a consistent formulation for these systems we need to construct  a 
positive definite $\eta $ such that
 inner product is well defined. 
In this section we intend to construct the positive definite $\eta $ explicitly
by using the spectral method as defined in Refs. \cite{spec,spec1}. 
The positive definite $\eta $ for  a $\eta $-pseudo Hermitian theory described
by the Hamiltonian $H$ is defined as \cite{spec,spec1},
\begin{equation}
{\bf \eta }= \sum_{i=1}^2|u_i><u_i|,
\label{5.1}
\end{equation}
where $|u_i>;\  \ i=1,2$ are the Dirac spinors associated with $H^\dagger ( = \eta H \eta ^{-1})$.
Following this method  we construct  the positive definite $\eta $ for both the
models  discussed above.

{\bf Case I: Pseudo-Hermitian Rasbha Interaction:}

\vspace{.1in}

In this case, $H^\dagger$ can be written from  Eq. (\ref{2.2}) as

\begin{equation}
H^\dagger = \left [
\begin{array}{cc} m_0 c^2 & (c+\lambda){\bf p_-}  \\
(c-\lambda){\bf p_+} & -m_0c^2 \end{array} \right ], \label{5.2}
\end{equation}
and let $\left ( \begin{array}{c} v_1 \\ v_2  \end{array}\right ) $ be the two component
spinor for $H^\dagger $ corresponding to the energy E, and satisfy,
\begin{equation}
H^\dagger \left ( \begin{array}{c} v_1 \\ v_2  \end{array}\right )
=E\left ( \begin{array}{c} v_1 \\ v_2  \end{array}\right )
\end{equation}
The spinors are calculated as
\begin{equation}
|u_1>=  \left ( \begin{array}{c} 1 \\ \frac{(c-\lambda)p_+}{E+m_0c^2} \end{array}\right );
\mbox{and }
|u_2>= \left ( \begin{array}{c} \frac{-(c+\lambda)p_-}{E+m_0c^2 }\\ 1  \end{array}\right )
\label{5.4}
\end{equation}
when $E$ is given in Eq. (\ref{2.5}) and $p_+, p_-$ are eigenvalues of the operator
{\bf $p_+$} and {\bf $p_-$} respectively.  
Substituting Eq. (\ref{5.4}) in  Eq. (\ref{5.1}) we obtain the positive definite
$\eta $, as
\begin{equation}
\eta_I = \left [ \begin{array}{cc} 1+\frac{(c+\lambda)^2}{(E-m_0c^2)^2}p_+p_- & \frac{2p_-(cE+
\lambda m_0c^2)}{E^2-m_0^2c^4} \\ \frac{2p_+(cE+
\lambda m_0c^2)}{E^2-m_0^2c^4} & 1+\frac{(c-\lambda)^2}{(E+m_0c^2)^2}p_+p_-
\end{array} \right ]
\end{equation}

{\bf Case II: Pseudo-Hermitian scalar interaction}\\

It is difficult to obtain the spinors associated with $H^\dagger $ in this case
for an arbitary $V(x)$ because $H^\dagger $ and the operators $p_{\pm}$ do not
have simaltaneous eigen functions as $[ H^\dagger, p_{\pm}]\neq 0$. However
for the special case when $V(x)$ is independent of x we can substitutes 
the operators $p_{\pm}$ by their eigenvalues as   $[ H^\dagger, p_{\pm}]= 0$.
We therefore construct the positive definite $\eta $ for the special case

For this model, 
\begin{equation}
H^\dagger = \left [
\begin{array}{cc} m_0 c^2 & c{\bf p_x}-V_0  \\
c{\bf p_x}+V_0 & -m_0c^2 \end{array} \right ]. \label{5.6}
\end{equation}
Let  $\left ( \begin{array}{c} v_1 \\ v_2  \end{array}\right ) $  be the two components
spinor for $H^\dagger $ corresponding to the energy E, and satisfy,
\begin{equation}
H^\dagger \left ( \begin{array}{c} v_1 \\ v_2  \end{array}\right )
=E\left ( \begin{array}{c} v_1 \\ v_2  \end{array}\right ),
\end{equation}
$|u_1>$ and $|u_2>$ are calculated as
\begin{equation}
|u_1>=  \left ( \begin{array}{c} 1 \\ \frac{cp_x+V_0}{E+m_0c^2} \end{array}\right );
\mbox{and }
|u_2>= \left ( \begin{array}{c} \frac{-(cp_x-V_0)}{E+m_0c^2 }\\ 1  \end{array}\right ),
\label{5.7}
\end{equation}
when $E$ is given by Eq. (\ref{3.7}).
Putting $|u_1>$ and $|u_2>$ from Eq. (\ref{5.7}) in Eq. (\ref{5.1}) we obtain 
\begin{equation}
\eta_{II} = \left [ \begin{array}{cc} 1+(\frac{cp_x+V_0}{E-m_0c^2})^2- & 
\frac{2Ecp_x-2V_0 m_0c^2}{E^2-m_0^2c^4}
 \\ \frac{2Ecp_x-2V_0 m_0c^2}{E^2-m_0^2c^4} &
 1+(\frac{cp_x-V_0}{(E+m_0c^2)})^2\end{array} \right ]
\end{equation}

We have constructed a positive definite $\eta $ for  both the examples discussed in this paper.
These positive definite $\eta $'s ensure the positive definite norms for  all the state vectors
associated with this theory and hence lead to a fullly consistent relativistic pseudo-Hermitian
theory.

\section{Conclusion}

\renewcommand{\theequation}{5.{\arabic{equation}}}
\setcounter{equation}{0}

 We have considered two completely
different pseudo-Hermitian interactions in relativistic quantum
mechanics. In the  first example, we have dealt  with Rashba interaction with imaginary
coupling which
plays an important role in studying spin Hall effect \cite{she,she11,she12,she13}.
 This interaction is shown to be parity pseudo-Hermitian and the
complete spectrum which we obtain by solving Dirac equation on the
plane is real. It will be interesting to see whether
the solutions of imaginary Rashba interaction lead to  any new
consequence in the study of spin Hall effect.  Scalar pseudo-Hermitian interaction has
 been constructed
with an arbitrary real parameter 
  in 1+1 dimension and is shown that the spectrum is
real  by solving corresponding Dirac equation. Using spectral method we have further
 constructed a positive definite metric operator, $\eta $ for both the
examples. Such positive definite $\eta $ ensures the positive definite norm 
for the state vectors. Thus we have a fully consistent quantum theory for  the
pseudo-Hermitian interactions in relativistic theory. 

\vspace{.2in}

{ \large \bf Acknowledgment:} We are grateful to Prof. Ali Mostafazadeh for
making valuable comments/suggestions on the earlier version of the manuscript
which have helped improve the paper substantially.

\end{document}